# Recommendations to overcome language barriers in the Vera C. Rubin Observatory Research Ecosystem

**José Antonio Alonso Pavón[1] Andrés Plazas Malagón[2]**

[1]**Faculty of Sciences, National Autonomous University of Mexico (UNAM), México City, Mexico,**
[2]**SLAC National Accelerator Laboratory, Kavli institute for Particle Astrophysics and Cosmology, Stanford University, Vera C. Rubin Observatory**

**ABSTRACT**

The report presents a comprehensive set of five recommendations to reduce language barriers within the Vera C. Rubin Observatory Research Ecosystem, promoting greater inclusion of researchers who are speakers of English as an additional language. Recognizing that English linguistic hegemony in science limits participation and productivity, the document proposes multilingual presentation formats, academic writing training, a Virtual Writing Center, language support programs, and writing retreats. Each recommendation is grounded in both pedagogical theory and empirical evidence, with an emphasis on collaborative, socially embedded approaches to scientific writing. The proposed academic writing training integrates constructivist and socio-cultural perspectives, emphasizing genre awareness, rhetorical competence, and reflective practices. The Virtual Writing Center would serve as a permanent infrastructure offering personalized tutoring and peer review support, while the language support programs address ongoing needs through workshops, consultations, and access to language tools. Writing retreats provide immersive environments for focused work and mentorship. The recommendations also encourage ethical use of AI tools for translation and writing assistance, fostering digital literacy alongside linguistic proficiency. Collectively, these initiatives aim to transform language from a barrier into a resource, recognizing multilingualism as an asset in global research collaboration. Rather than offering a one-size-fits-all solution, the document advocates for adaptable, community-driven strategies that can evolve within the diverse institutional and disciplinary contexts of the Rubin Research Ecosystem. By implementing these practices, the Ecosystem could lead efforts to democratize scientific communication and foster a more equitable, multilingual research culture.

## Full Abstract

Scientific communication has been largely conducted in English as the main language, with often little accommodations for linguistic diversity. However, as global participation in research expands and accessibility improves, new opportunities for multilingual engagement are emerging, helping to bridge language gaps and foster more inclusive and productive collaborations.

This document provides a set of recommendations to mitigate these challenges within the Vera C. Rubin Observatory Research Ecosystem—defined broadly here as the Rubin Observatory Construction and Operations staff, members of the Rubin Legacy Survey of Space and Time (LSST) Science Collaborations, the LSST Discovery Alliance, Rubin funding agencies, and the global research community—and beyond.

The recommendations aim to foster a more inclusive scientific environment by suggesting initiatives such as enabling multilingual presentations and materials with translation support, providing tailored academic writing training to enhance science communication, and establishing a Virtual Writing Center. This center could offer personalized consultations, workshops, and peer review sessions, leveraging existing technologies and a tutor network to support researchers across the Rubin Ecosystem.

Additional suggestions include the organization of writing retreats or workshops (or dedicated sessions in meetings) to foster dedicated spaces focused on academic writing in English as a Second Language, and the creations language support programs designed to address specific challenges faced by speakers of English as an additional language.

The recommendations in this report are based on evidence from the literature and best practices from other initiatives while considering the specific needs and constraints of the Rubin Observatory Ecosystem.

By publishing this document, the recommendations are expected to potentially benefit not only Rubin but also other scientific collaborations and institutions. This work underscores the importance of inclusivity and accessibility in science, providing a roadmap to reduce language barriers and enhance the participation and productivity of researchers from diverse linguistic backgrounds.

This work was funded by the LSST Discovery Alliance (LSST-DA) via the proposal **"Recommendations to overcome language barriers in the Rubin Observatory"**, awarded as part of the "Inclusive Collaboration: Policies and Best Practices (2023)" call ([LSST Discovery Alliance, 2023](#)). It constitutes a set of recommendations to be considered by the members of the Vera C. Rubin Observatory research ecosystem and the astronomical and scientific community in general, and it is not an official Rubin Observatory document.

## Executive Summary

### Recommendation 1: Multilingual Presentations and Materials.

Enable speakers of English as an additional language to present in their first languages at the events hosted within the Rubin Research Ecosystem (RRE), supported by translated materials or live interpretation.

### Recommendation 2: Academic Writing Training

Implement formal academic writing training specifically tailored to science (astronomy).

### Recommendation 3: Virtual Writing Center

Create a Virtual Writing Center to improve academic writing skills. The Center would offer personalized consultations, exchange meetings and seminars, leveraging technologies like Zoom, Microsoft Teams and Slack to serve a dispersed scientific community. The model is based on tutors selected from the Rubin Research Ecosystem itself, who should receive psycho-pedagogical training to effectively guide students and researchers in their specific writing needs.

### Recommendation 4: Language Support Programs

Implement language support programs additional to the Writing Center such as one-on-one writing consultations, group workshops, peer review sessions, and feedback sessions specifically tailored for speakers of English as an additional language, highlighting the linguistic challenges and potential solutions in academic research.

### Recommendation 5: Writing Retreats and Workshops

Implement Writing Retreats designed to provide scholars with dedicated, uninterrupted writing time, create a supportive and distraction-free environment, and that offer personalized mentorship for researchers, especially speakers of English as an additional language.

## Introduction

The Vera C. Rubin Observatory, situated on Cerro Pachón in Chile, is a groundbreaking astronomical facility set to conduct the **Legacy Survey of Space and Time** (LSST; Ivezić, et al., 2019) starting in 2025. Over a decade, this observatory will use the 8.4-meter Simonyi Survey Telescope, a 3200-megapixel camera (the LSSTCam), and optical and near-infrared filters to comprehensively map the southern celestial hemisphere. Its capabilities include imaging the entire visible sky every 3-4 nights, potentially discovering approximately 17 billion stars and 20 billion galaxies (Rubin Observatory, n.d.). The observatory aims to advance scientific understanding in key areas such as dark matter, dark energy, the solar system, the Milky Way, and transient science. The Rubin Observatory is a complex, globally distributed organization with hundreds of employees working across multiple international sites. Its construction and operations involve diverse teams and subsystems, and the broader Rubin community also encompasses thousands of researchers organized into eight LSST Science Collaborations. In addition, the community includes the LSST Discovery Alliance, an independent non-profit organization dedicated to maximizing the scientific impact of the LSST by facilitating a diverse global network of scientists to access and analyze the vast amount of data generated by the survey, promoting inclusive participation and interdisciplinary collaboration in astronomical research. Altogether, the Rubin Observatory Construction and Operations staff, the LSST Science Collaborations, the LSST Discovery Alliance, the Rubin funding agencies, and the global research community will be referred to in this document as the **Rubin Research Ecosystem (RRE)**.

Effective communication of research findings is crucial in science, especially given the global nature of modern scientific endeavors. (Nat Rev Bioeng Editorial, 2023). The dominance of English in scientific communication can significantly limit the full participation of speakers of English as an additional language in the global scientific community (Ferguson, et al., 2011; Lillis & Curry, 2010). Strategies supporting speakers of English as an additional language can address systemic barriers to equity, diversity, and inclusion, and enhance international scientific collaborations, which are particularly vital in fields like astronomy.

Many researchers struggle with academic communication and academic writing, even in their first language (Gastel & Day, 2022). The challenges are amplified when writing in a foreign language, necessitating comprehensive approaches to improve scientific writing skills and communication of research findings. Researchers in science, technology, engineering and mathematics (STEM) fields, and especially speakers of English as an additional language, encounter significant challenges when writing academic papers in English. Despite being fluent, they often find scientific writing more difficult, anxiety-inducing, and less satisfying compared to writing in their first language (Hanauer & Englander, 2013; Hanauer, Sheridan & Englander, 2019). Early-career scientists who are not speakers of English as a first language face significant challenges in scientific communication, including increased difficulty reading literature, writing papers, and participating in international conferences (Ferguson, et al., 2011; Wallwork, 2016). These challenges directly impact scientific communication, causing longer preparation times, increased revision needs, and reduced confidence in research submissions, potentially affecting overall professional performance (Ferguson, et al., 2011; Hanauer & Englander, 2011).

This document provides a set of five recommendations to address critical challenges within the Vera C. Rubin Observatory Research Ecosystem and the broader academic research community. It explores the challenges, strengths, and potential areas for improvement among an international cohort of researchers representing diverse backgrounds, career stages, and disciplinary perspectives. The approach is based on Vygotsky's socio-constructivist view of writing as a socially mediated and collaborative process (Vygotsky, 1978), but it also incorporates critical viewpoints in English for Research Publication Purposes (ERPP). Following the critical pedagogies of Englander and Corcoran (2019), we recognize that writing in English for publication is not a neutral act, but rather one molded by structural power dynamics that favor some languages, epistemologies, and geographies over others. Thus, training and support systems must go beyond technical proficiency to involve researchers in critical reflection on language hegemony, epistemic inclusivity, and the politics of scientific legitimacy. Our goal with this set of recommendations is to promote not just more effective communication, but also a more just and multilingual research environment. These recommendations are intended as a flexible set of options rather than a prescriptive plan—different parts of the RRE may choose to implement one, several, or all of them, depending on their specific needs and capacities.

## Recommendations

To overcome language barriers in science, an approach addressing linguistic complexities is essential. By creating an equitable environment for speakers of English as an additional language, institutions can develop diverse talent and perspectives, ultimately accelerating scientific progress globally. These recommendations aim to empower researchers by recognizing language as a dynamic communication tool, not a rigid barrier, and prioritize scientific clarity and authentic expression.

These recommendations emphasize written communication as the primary vehicle for academic advancement, given its foundational role in publishing and professional visibility. While oral skills remain essential, they typically follow written outputs. The document therefore prioritizes writing without dismissing the importance of effective verbal communication in the research ecosystem. This approach provides a balanced perspective on scientific communication, offering support through tailored interventions that enhance research productivity and knowledge dissemination.

## Recommendation 1: Multilingual Presentations and Materials.

Speakers of English as an additional language face substantial research communication challenges, often investing significant additional effort in multilingual communication. Breaking down these language barriers can unlock research potential and enhance global scientific innovation (Amano, et al., 2023). Multilingualism is crucial for promoting academic diversity and ensuring high-quality research transcends linguistic boundaries (Kulczycki, et al., 2020; Pölönen & Kulczycki, 2023).

To challenge English linguistic hegemony in scientific communication, the Rubin Research Ecosystem could allow researchers to present in their first languages at in-person or hybrid meetings, fostering a multicultural environment and raising awareness about language privilege. This approach helps speakers of English as an additional language express ideas more naturally while exposing speakers of English as a first language to multilingual communication challenges (Amano, et al., 2023). While this initiative primarily aims to empower speakers of English as an additional language, it is equally valuable to invite speakers of English as a first language—particularly those with multilingual proficiency—to participate in these forums. Doing so not only reinforces the multilingual ethos of the Rubin Research Ecosystem but also allows English-dominant researchers to gain exposure to alternative rhetorical structures and communication styles that vary across linguistic and cultural contexts (Pérez-Llantada, 2012; Taylor & Chen, 1991). This experience may encourage broader participation in non-English scientific spaces and deepen mutual understanding within the international research community.

To facilitate multilingual communication for the RRE members, translation infrastructure combining Artificial Intelligence (AI) technologies and human expertise could be deployed at events of the Rubin community (e.g., Rubin meetings such as the annual Rubin Community Workshop, collaboration meetings from the LSST Science Collaborations, workshops, sprint days, etc), including professional interpreters, AI translation platforms, and real-time translation tools to enable seamless cross-linguistic research collaboration (Steigerwald, et al., 2022; Anon., 1954).

Translation platforms offer comprehensive language support across 60+ languages. These tools provide flexible communication formats for in-person, virtual, and hybrid conferences, integrating advanced AI technologies with professional human interpretation. Precise participant-specific quotes are crucial for effective budgeting and understanding scalability options. Platforms like Boostlingo[1], Wordly[2], KUDO[3], or Interprefy[4], which offer comprehensive language solutions by providing remote interpreting, AI speech translation, and live captions for various event types, might be useful to cater RRE events by offering comprehensive language support that addresses multiple critical communication needs. While these platforms offer promising solutions by combining AI technologies with optional human interpretation, their capacity to accurately handle highly specialized scientific terminology remains a legitimate question. Some of these platforms—particularly KUDO and Interprefy—indicate use in academic and professional research settings such as university courses, webinars, and large-scale multilingual events. However, publicly available information does not always specify whether their deployment has included domain-specific scientific conferences requiring high-precision terminology. As such, while their infrastructure appears adaptable to research contexts, further evaluation would be necessary to assess their performance in highly technical scientific communication.

To test feasibility and gather feedback, multilingual presentation support could be piloted at one major Rubin-hosted event during the first year—such as the Rubin Community Workshop or a Science Collaboration meeting—with the aim of expanding support to additional events annually if demand and resources allow.

## Recommendation 2: Academic Writing Training

Academic Writing training should focus on scientific writing conventions, incorporating both genre-specific instruction and an iterative writing process ([Schickore, 2008](#); [Miller, 1984](#)). Contextual learning is crucial, emphasizing that effective communication transcends literal translation. Researchers must develop a nuanced understanding of academic discourse, mastering not just vocabulary, but linguistic structures, rhetorical conventions, and unspoken communication norms. By cultivating this deeper linguistic awareness, researchers can transform language from a barrier into a bridge of intellectual connection.

Designing effective Academic Writing training will require building a solid pedagogical framework. A pedagogical framework provides a strategic blueprint for educational design, synthesizing diverse philosophical perspectives to develop effective training programs ([Brown, 2006](#)). Proficient academic writing requires integrating textual, contextual, and linguistic knowledge. Consequently, training programs should prioritize developing participants' autonomy in manuscript revision and self-editing, emphasizing reflective practices and collaborative learning through shared writing experiences.

The strategy suggested here includes rooting the pedagogical framework in Lev Vygotsky's Social Development Theory and Constructivist theory. Vygotsky's insights reframe writing as a socially embedded practice, creating a collaborative learning environment where writing skills emerge through peer support and shared knowledge. By emphasizing social interactions within the scientific community, writing transforms from an individual skill to a collective knowledge construction process ([Vygotsky, 1978](#); [Liu & Matthews, 2005](#)). Constructivism could complement and further enrich this training design by conceptualizing knowledge as an active, collaborative construction. This approach repositions educators as architects of meaningful learning experiences, fundamentally reshaping how academic writing is understood and taught. Rather than viewing writing as a fixed skill, this perspective presents it as a dynamic process of meaning-making ([Gergen, 1985](#); [Fox, 2001](#); [Gordon, 2009](#)).

This training can be supported by [Englander and Corcoran's (2019)](#) critical plurilingual pedagogy, whose emphasis is on empowering researchers through contextualized genre awareness, collaborative learning environments, and the strategic use of all accessible linguistic resources both internally and externally. In accordance with their report on the work done in the University of Mexico, the training program development could incorporate local knowledge from researchers currently working on the RRE, respect current multilingual practices, and assist participants in navigating publishing in English without losing their linguistic, disciplinary or epistemic identities (*i.e.* Astronomy). Since the RRE is multinational and diverse, this approach is particularly important. This recommendation proposes structured academic writing training as a foundational intervention. It emphasizes the development of genre awareness, rhetorical strategies, and disciplinary conventions in scientific English. Unlike ad-hoc workshops or mentoring formats, this training is designed as a formal instructional program, potentially delivered through short courses or intensive modules.

To ensure training effectiveness, institutions within the RRE could create **interdisciplinary teams** with language experts and subject matter specialists from various scientific fields. This collaborative approach provides an integrative learning experience that addresses both linguistic and content-specific challenges (Hanauer & Englander, 2011). By providing ongoing language support, researchers can effectively bridge communication gaps.

Training should include strategies such as the development of comprehensive resources like style guides, academic phrase-banks, vocabulary lists, and citation tools to help speakers of English as an additional language navigate scientific writing complexities. Additionally, institutions are encouraged to follow an approach that promotes self-editing skills, collaborative learning strategies, and technological support to empower researchers in improving their academic writing independently.

Prioritizing collaborative learning strategies is highly recommended. Approaches like peer review sessions, group discussions about reading and writing topics, and team-based learning for analyzing and improving scientific papers, not only enhance learning but also mimic the collaborative nature of scientific research itself, preparing researchers for the types of interactions they will encounter in their professional lives.

As part of this training, participants should also engage with the capabilities and limitations of emerging AI writing tools (*e.g.* ChatGPT[5]) translation tools (*e.g.* DeepL[6]), as well as bibliographic reference gathering, using tools (*e.g.* ResearchRabbit[7], Elicit[8]) to discover references beyond conventional search methods (Del Giglio & Pereira da Costa, 2023). AI tools offer valuable support in the scientific writing process, particularly for tasks such as brainstorming, drafting titles and abstracts, translating, and revising texts. When used strategically—especially after core sections like methods and results have been manually developed—they can enhance clarity and structure. However, reliance on these tools also raises concerns around authorship, data confidentiality, citation ethics, and the erosion of writing proficiency. Integrating critical AI literacy into academic writing training will help researchers use these technologies as complementary aids, rather than substitutes for human judgment, fostering both effective and ethically sound practices. (Del Giglio & Pereira da Costa, 2023).

To ensure continuity and sustained impact, academic writing training could be offered on an annual or biannual basis, depending on demand and availability of facilitators. Shorter sessions, such as thematic webinars or refresher modules, could be scheduled quarterly to support ongoing skill development and community engagement.

In parallel with training for early-career researchers, programs should also consider mentoring capacity-building for supervisors and senior researchers. While many have deep expertise in scientific writing, they may not always have the vocabulary or pedagogical strategies to effectively mentor others—particularly those writing in English as an additional language. Drawing on applied linguistics literature (*e.g.*, [Swales & Feak, 2012](#); [Cargill & O'Connor, 2021](#)), short workshops or guides can help experienced researchers become more confident and explicit writing mentors within their labs and teams. This approach supports the emergence of sustainable, localized writing cultures grounded in disciplinary expertise.

## Recommendation 3: Virtual Writing Center

Writing centers are dynamic academic environments dedicated to cultivating advanced writing skills across higher education, strategically developing students' intellectual capabilities beyond traditional grammatical instruction. Led by a director and trained tutors, these centers offer individualized 30–60-minute sessions addressing various writing needs, from essay planning to research paper revision. Beyond one-on-one tutoring, writing centers conduct workshops and training programs designed to develop critical thinking, analytical skills, and effective communication ([López Gil (coord), 2013](#); [McKinley, 2010](#)). These are communication strategies that prepare users for complex professional challenges in today's rapidly evolving job market ([Mckay & Simpson, 2013](#)).

Writing centers promote educational equity by supporting users from diverse backgrounds, including speakers of English as an additional language. They advance academic literacy by helping users develop critical skills in formulating arguments, synthesizing information, and adapting writing for different audiences' specific needs ([McKinley, 2010](#); [Mckay & Simpson, 2013](#); [Ginting & Barella, 2022](#)). By fostering a supportive environment that emphasizes revision and ongoing improvement, writing centers help students become confident, self-sufficient writers prepared for academic and professional challenges. Far from being simple tutoring spaces, writing centers are truly transformative spaces that cultivate academic excellence, professional growth, and social inclusion.

A Virtual Writing Center could be created as a digital initiative to provide accessible writing services to a geographically dispersed scientific community. Using virtual communication platforms the center could offer personalized writing consultations, idea exchange meetings, and peer review sessions. Focused on academic writing in astronomy and physics, the center could aim to improve scientific publication quality and communication abilities for its members. The approach involves gradually identifying skilled RRE members to undergo initial academic writing training and become tutors, with an emphasis on specialized training that includes both technical writing skills and psycho-pedagogical techniques to effectively support diverse community members. While [Recommendation 2](#) offers initial academic writing instruction, the Writing Center provides long-term, flexible support beyond the classroom. Its purpose is to serve as a sustained institutional infrastructure where researchers can access ongoing, personalized help with their writing projects, participate in peer feedback, and engage in informal writing exchanges and seminars throughout their career stages.

A weekly drop-in writing session, based on the popular '*Shut Up and Write!*' format—a structured approach where participants meet (virtually or in person) for short, focused writing intervals followed by brief breaks—could be organized by the Virtual Writing Center in addition to individual consultations and asynchronous support. This would promote productivity and a sense of writing community. In addition to providing a common area for accountability and support, these low-barrier, community-focused meetings assist researchers in maintaining steady progress on their writing projects. Additionally, this ongoing involvement prepares the basis for more intensive writing sessions, such the retreats outlined in [Recommendation 5](#).

The Virtual Writing Center can also serve as an ongoing space for researchers to receive guidance on the responsible use of AI tools in academic writing, offering curated resources, workshops, and peer discussions on best practices for integrating AI-assisted drafting, grammar correction, and translation while maintaining academic integrity. Dr. May Chiao, former chief editor of *Nature Astronomy*, recently highlighted at ESOGPT24 both the potential and risks of AI in scientific publishing. She emphasized that while AI can assist with tasks like copy-editing and structuring abstracts or titles, it cannot replace human creativity, authorship, or accountability. Springer Nature is now developing specialized AI tools and implementing detectors to identify AI-generated content ([Chiao, 2024](#)). Research on AI-mediated communication also shows that disclosing the use of tools like machine translation can reduce perceived authenticity and trustworthiness—though partial use is generally received more positively than full reliance ([Glikson, 2024](#)). These findings reinforce the need for critical literacy, transparency, and strategic use of AI within the Rubin Research Ecosystem.

The proposal for the Rubin Research Ecosystem's Virtual Writing Center emphasizes a cost-effective, community-driven approach to improving academic writing skills. Initially, there would be an investment in training and personnel, but the long-term goal is to create a self-sustainable model where the community guides its own development. By leveraging collective experience and knowledge within the ecosystem, the center aims to minimize ongoing operational expenses while continuously improving academic writing support through a cascading mentorship approach that reduces dependence on external resources. After an initial six-month development and tutor training phase, the Center could begin offering services such as personalized writing consultations, group seminars, and peer review sessions on a rolling basis throughout the year. A minimum cadence of one seminar or exchange session per quarter is suggested to ensure sustained participation and engagement.

## Recommendation 4: Language Support Programs

Speakers of English as an additional language encounter substantial challenges in mastering academic English, which can significantly impede research productivity and knowledge dissemination. To effectively navigate these linguistic barriers, a multifaceted approach is crucial—one that integrates reading, writing, and oral skills as interdependent components of academic communication. Language support programs should promote consistent reading practices that progress from accessible to complex texts, thereby reinforcing both vocabulary acquisition and rhetorical awareness. These reading strategies should be complemented by interactive tools (e.g., flashcards), writing consultations, and discussion-based activities that help participants articulate scientific concepts across written and spoken modalities. This integrated model not only enhances individual skill sets but also prepares researchers for the diverse communication demands of academic life. This approach involves comprehensively exploring academic sentence structures, nuanced idiomatic expressions, and the sophisticated linguistic conventions that govern scholarly discourse ([Marta & Ursa, 2015](); [Durgumahanthi, 2024]()).

A strategic approach to language support would encompass diverse collaborative learning opportunities designed to enhance academic research communication. These targeted interventions—including one-on-one writing consultations, group workshops, peer review sessions, and specialized feedback sessions—create a robust ecosystem for linguistic and scholarly development. This collaborative model not only supports individual growth but also fosters a supportive, intellectually dynamic environment where researchers can learn from each other's diverse perspectives and writing styles. The goal would be to create a collaborative learning environment that supports the diverse communication needs of the RRE.

The program would demand careful planning, including comprehensive guidelines, rigorous participant training in constructive feedback techniques, and the cultivation of a respectful, collaborative learning environment. By strategically integrating advanced technological platforms, the initiative can enhance operational efficiency and create multiple review stages that provide diverse, nuanced perspectives on academic writing. The overarching objective extends beyond mere writing or oral communication skills development; it aims to foster a culture of collaborative learning, and comprehensively prepare scholars for the complex linguistic challenges inherent in academic and professional writing contexts.

The implementation would follow a carefully structured two-phase approach. The initial six-month period could focus on establishing foundational infrastructure, including developing sophisticated digital collaboration platforms, crafting detailed review guidelines, and designing targeted workshops that address critical language learning strategies. Following deployment, services such as one-on-one consultations, peer sessions, and group workshops could be offered monthly or bimonthly. Program evaluation should occur every six months to assess outcomes and inform adjustments in frequency, content, and delivery methods.

In addition to structured consultations and workshops, Language Support Programs could offer on-demand micro-support, such as weekly "language office hours" or asynchronous Q&A spaces, where researchers can ask quick questions about usage, grammar, or phrasing—such as "participate in" versus "participate on," or how to refer to "Figure 2." Additionally, training should introduce corpus-based search strategies using tools like *SkELL*[9] or *Linggle*[10], which allow users to explore frequency-based usage patterns and make more confident, autonomous language decisions in real time.

These programs offer recurring, low-cost opportunities for language development, including one-on-one consultations, peer review sessions, and group workshops. Unlike the Virtual Writing Center, which would serve as a centralized system, Language Support Programs can be implemented independently by smaller teams or projects within the RRE and can be adapted to specific disciplinary or institutional needs.

Success could be evaluated through a multifaceted assessment framework that examines tutor development, community engagement, writing output quality, participant satisfaction, and the institution's growing self-sufficiency in academic writing support. This nuanced approach would ensure a systematic and holistic understanding of the program's transformative potential, meticulously tracking both quantitative metrics and qualitative insights to comprehensively measure the initiative's impact across multiple dimensions.

## Recommendation 5: Writing Retreats and Workshops

Writing retreats are structured academic workshops that provide scholars dedicated, uninterrupted time to focus on writing projects. Varying from 2-3 day facilitated events to non-facilitated sessions, these retreats include team-building exercises, manuscript preparation, and collaborative discussions. These retreats provide a distraction-free environment that enables researchers to focus intensely on manuscript development while receiving targeted mentorship and peer support. Retreats create a community of practice that brings together writers of different experience levels, encouraging knowledge sharing and constructive feedback across disciplines (Kramer & Libhaber, 2016; Kornhaber, et al., 2016).

Speakers of English as an additional language face significant linguistic challenges that require extraordinary mental resilience, potentially impeding research productivity (Durgumahanthi, 2024). Hence, writing retreats offer an inclusive solution, providing a supportive environment where researchers can overcome language barriers and develop their academic writing skills.

Writing retreats offer more than stress relief; they provide a supportive environment that helps researchers, especially speakers of English as an additional language, overcome linguistic insecurities and build confidence in academic writing (Stevenson, 2021). By creating a sanctuary-like atmosphere, these retreats enable participants to transform their perception of language barriers, viewing their multilingual background as a strength rather than a limitation. The supportive setting allows researchers to reframe their approach to academic communication, ultimately empowering them to effectively convey complex scientific ideas across linguistic boundaries.

Writing retreats differ from all previous recommendations by providing immersive, in-person environments designed for deep writing engagement over several consecutive days. They allow scholars to work on advanced projects with sustained mentoring and are especially helpful for overcoming writing fatigue or writer's block.

Writing retreats could be held once or twice per year, ideally during low-demand periods in the Rubin calendar. Standalone workshops—focused on manuscript revision, publication strategies, or oral communication—could be offered more frequently (e.g., quarterly), and may be hosted virtually to broaden access and reduce costs. These in-person retreats can be further enhanced by virtual pre-retreat writing communities (*e.g.*, weekly writing sessions through the Virtual Writing Center), which allow participants to prepare in a supportive, low-pressure environment and arrive with clearer goals and active drafts in progress.

By offering personalized mentorship, professional resources, and a collaborative atmosphere, these retreats aim to transform academic writing from a challenging task into an achievable, supportive process. The comprehensive approach focuses not just on immediate productivity, but on building long-term writing skills and confidence among researchers, particularly those facing linguistic or professional challenges in scholarly communication.

## Summary and Concluding Remarks

The [first recommendation](#) focuses on creating multilingual presentations and materials to enhance **accessibility for speakers of English as an additional language** in scientific communication. By enabling researchers to present in their first languages with translation support, the RRE aims to democratize academic discourse and challenge linguistic barriers. The approach proposes advanced translation infrastructure combining AI technologies and human expertise to create a more equitable scientific collaboration environment. The biggest challenge that could face the RRE if it decides to implement this recommendation would be the financial burden associated to the simultaneous translation technologies required to offer a multilingual setting for its members. The costs may vary significantly depending on the provider of the services, but it is also expected that with the advancement of AI translation technologies those prices would drop eventually, making it more affordable for the RRE.

Recommendations 2 to 5 are all designed to enhance academic communication skills, particularly for speakers of English as an additional language. While they share a common goal, they approach it through distinct strategies: [Recommendation 2](#) emphasizes foundational academic writing training and pedagogical design; [Recommendation 3](#) focuses on building a permanent support infrastructure through a Virtual Writing Center; [Recommendation 4](#) addresses ongoing and flexible support mechanisms through modular programs; and [Recommendation 5](#) proposes immersive, time-bound retreats for concentrated writing development. Each offers a different scope, format, and level of engagement, which together create a layered ecosystem of language support within the Rubin Research Ecosystem.

The [second recommendation](#) suggests implementing specialized academic writing training for the RRE's members. This training should aim to address the unique challenges faced by speakers of English as a first and as an additional language in scientific writing by providing **discipline-specific instruction** that combines genre and process approaches to writing. The recommendation emphasizes creating interdisciplinary teaching teams with language experts and subject matter specialists, training for clear English writing proficiency skills, and developing comprehensive resources like style guides and academic phrase banks. For this recommendation, the greatest challenge would be creating an initial team of instructors that can offer specialized genre and process-based approaches to academic writing. Once a generation of the RRE's members is trained in the details around this specific scientific practice, they can provide the ensuing training for future generations.

The [third recommendation](#) advises establishing the RRE Writing Center as a virtual initiative designed to enhance academic writing skills. By leveraging technologies like Zoom and Slack, the center could provide personalized consultations, exchange meetings, and seminars to a dispersed scientific community. This model would recruit tutors from within the RRE, who would have already received specialized psycho-pedagogical training to effectively guide speakers of English as an additional language and first language. Given the practicality of online communications and the international nature of the Ecosystem, courses given virtually would reduce the financial burden of travel expenses for participants, trainers and the ecosystem. With this gradual and self-sustaining approach, the center aims to elevate the quality of scientific publications and strengthen the research community's collaborative capabilities.

The Language Support Programs suggested in the [fourth recommendation](#) focus on creating a supporting set of tasks that could supplement the writing training to offer a follow-up that addresses the ongoing linguistic challenges faced by native and speakers of English as a first and as an additional language The program aims to improve English language comprehension through strategic approaches such as consistent reading practice, building discipline-specific vocabulary, and contextual learning. The strategy suggested this recommendation includes one-on-one writing consultations, group workshops, peer review sessions, and feedback mechanisms. These activities are designed to create a collaborative learning environment within the RRE, potentially evolving into a dedicated Writing Center -which constitutes the core of the fourth recommendation. The goal is to support diverse communication needs, enhance research productivity, and help members of the RRE effectively share their scientific findings across linguistic boundaries. By implementing these targeted language support programs, the initiative seeks to democratize academic discourse and provide equitable opportunities for researchers to communicate their work effectively.

The [last recommendation](#) proposes implementing writing retreats as a strategic approach to support scholars in producing scholarly publications amid challenging workloads. These retreats provide a dedicated, distraction-free environment where researchers can intensively focus on writing, offering mentorship and structured support. By bringing together academics, the retreats could foster a collaborative community that helps participants overcome writing barriers, develop academic writing skills, and boost confidence in other language-related areas such as speaking. The retreats aim to address time constraints and linguistic challenges faced by researchers, particularly speakers of English as an additional language. By creating a supportive atmosphere with experienced facilitators, the writing retreats seek to alleviate writing-related anxiety and provide participants with practical strategies for maintaining writing momentum beyond the retreat itself.

The structural recommendations recognize that academic communication is not a one-size-fits-all endeavor. Different research types - from scientific articles to review papers - require nuanced approaches. This flexibility allows researchers to adapt their communication strategies while maintaining rigorous intellectual standards. The ultimate contribution of these recommendations lies in their commitment to **elevating academic communication**. By providing clear guidelines that are simultaneously structured and adaptable, researchers are equipped to produce more precise, meaningful, and accessible scholarly work. The approach signals a sophisticated understanding of academic writing as a dynamic, context-sensitive form of intellectual exchange.

The recommendations presented in this document present an approach to addressing the complex linguistic challenges inherent in scientific communication. By developing a strategy that encompasses multilingual support, specialized academic writing training, targeted language support programs, a virtual writing center, and immersive writing retreats, the Rubin Research Ecosystem can create a more inclusive, supportive, and intellectually dynamic environment for researchers across diverse linguistic backgrounds.

These strategies transcend traditional language correction methodologies, instead focusing on empowering scientists to become confident, articulate communicators of their research. For instance, the multilingual presentation support could enable a Brazilian astrophysicist to present groundbreaking research findings in Portuguese, with real-time translation ensuring that the nuanced scientific insights are not lost in linguistic translation. Similarly, the virtual writing center could provide personalized mentorship to a Korean researcher struggling to articulate complex theoretical concepts in English, helping them refine their manuscript to meet international publication standards.

By recognizing language not as a barrier but as a rich, complex medium of intellectual exchange, these recommendations aim to democratize scientific discourse. They acknowledge that brilliant scientific insights can emerge from researchers worldwide, regardless of their first language. The proposed approaches—from AI-powered translation technologies to peer review workshops—are designed to level the academic playing field, ensuring that linguistic background does not determine the visibility or impact of scientific research.

Moreover, these strategies have profound implications beyond individual research productivity. By fostering a more inclusive research environment, the Rubin Research Ecosystem can catalyze unprecedented cross-cultural collaborations, accelerate knowledge sharing, and drive global scientific innovation. When researchers from diverse linguistic and cultural backgrounds can communicate seamlessly, the potential for groundbreaking interdisciplinary discoveries increases exponentially.

Ultimately, breaking down linguistic barriers is about more than improving communication skills—it's about creating a truly global, equitable scientific community. These recommendations represent a bold commitment to recognizing and celebrating linguistic diversity as a strength, not a limitation. By providing researchers with the tools, support, and confidence to share their scientific narratives effectively, we can unlock a more interconnected, innovative, and intellectually vibrant research landscape that transcends traditional linguistic boundaries and propels scientific understanding forward.

## Footnotes

1. https://boostlingo.com ↩

2. https://www.wordly.ai ↩

3. https://kudo.ai/ ↩

4. https://www.interprefy.com/ ↩

5. https://chatgpt.com ↩

6. https://www.deepl.com ↩

7. https://www.researchrabbit.ai/ ↩

8. https://elicit.com/ ↩

9. https://skell.sketchengine.eu ↩

10. https://search.linggle.com ↩